\documentclass{cjaa}                    

\usepackage{graphicx}                   
\input{epsf.sty}                        
\input{psfig.sty}                       
\usepackage{amssymb,natbib}

\def\cmmoinsdeux{\mbox{ cm}^{-2}}

\def\microns{\mbox{ } \mu \mbox{m}}

\def\Msol{\mbox{ }M_{\odot}}
\def\Rsol{\mbox{ }R_{\odot}}

\def\mags{\mbox{ magnitudes}}

\def\adeg{^{\circ}}

\def\amin{^\prime}

\def\nh{N_{\rm H}}

\def\ltsima{\; \buildrel < \over \sim \;}
\def\simlt{\lower.5ex\hbox{\ltsima}}            
\def\gtsima{\; \buildrel > \over \sim \;}
\def\simgt{\lower.5ex\hbox{\gtsima}}            

\begin{document}

   \title{Obscured sources and Supergiant Fast X-ray Transients: new
classes of high mass X-ray binaries 
\footnotetext{Based on
observations collected at the European Southern Observatory, Chile
(proposals ESO N$\adeg$ 070.D-0340, 071.D-0073, 073.D-0339, 075.D-0773
and 077.D-0721).}  
}


   \author{Sylvain Chaty
\mailto{chaty@cea.fr}
   }
   \offprints{S. Chaty}                   

   \institute{Laboratoire AIM, CEA/DSM - CNRS - Universit\'e Paris Diderot,
DAPNIA/Service d'Astrophysique, B\^at. 709, CEA-Saclay,
FR-91191 Gif-sur-Yvette Cedex, France
\\
             \email{chaty@cea.fr}
        }

   \date{Received \today; accepted \today}

   \abstract{A new type of high-energy binary systems has been
revealed by the {\it INTEGRAL} satellite.  These sources are in the
course of being unveiled by means of multi-wavelength optical, near-
and mid-infrared observations. Among these sources, two distinct
classes are appearing: the first one is constituted of intrinsically
obscured high-energy sources, of which IGR~J16318-4848 seems to be the
most extreme example. The second one is populated by the so-called
supergiant fast X-ray transients, with IGR~J17544-2619 being the
archetype. We report here on multi-wavelength optical to mid-infrared
observations of a sample of 21 {\it INTEGRAL} sources. We show that in
the case of the obscured sources our observations suggest the presence
of absorbing material (dust and/or cold gas) enshrouding the whole
binary system.  We finally discuss the nature of these two different
types of sources, in the context of high energy binary systems.
 }

   \authorrunning{S. Chaty}            
   \titlerunning{Obscured sources and SFXTs: new classes of HMXBs}  


   \maketitle
%
%
\section{Introduction}

The {\it INTEGRAL} observatory has performed a detailed survey of the
galactic plane and the ISGRI detector on the IBIS imager has
discovered many new high energy sources, most of all reported in
\cite{bird:2007} (see also {\em
http://isdc.unige.ch/$\sim$rodrigue/html/igrsources\-.html}).  
The most important result of {\it INTEGRAL} to date
is the discovery of many new high energy sources --concentrated in the
Galactic plane, and in the Norma arm (see
e.g. \citeauthor{chaty:2005a} \citeyear{chaty:2005a})--, exhibiting
common characteristics which previously had rarely been seen. Many of
them are high mass X-ray binaries (HMXBs) hosting a neutron star
orbiting around an O/B companion, in most cases a supergiant
star. They divide into two classes: some of the new sources are very
obscured, and exhibiting a huge intrinsic and local extinction,
and the others are HMXBs hosting a
supergiant star and exhibiting fast and transient outbursts: an unusual
characteristic among HMXBs: they are therefore called Supergiant Fast
X-ray Transients (SFXTs, 
\citeauthor{negueruela:2006a} \citeyear{negueruela:2006a};
\citeauthor{sguera:2005} \citeyear{sguera:2005}).
%
High-energy observations are not sufficient to reveal the nature of
the newly discovered sources, since the {\it INTEGRAL} localisation
($\sim 2\amin$) is not accurate enough to unambiguously pinpoint the
source at other wavelengths. Once X-ray satellites such as {\it
XMM-Newton}, {\it Chandra} or {\it Swift} provide an arcsecond
position, the hunt for the optical counterpart of the source is open.
However, the high level of absorption towards the galactic plane makes
the near-infrared (NIR) domain more efficient to identify these
sources.  We first report on multi-wavelength observations of two
sources belonging to each class described above, then give general
results on {\it INTEGRAL} sources, before
discussing them and concluding.

\section{Observations and results} \label{observations}

The multiwavelength observations described here were performed at the
European Southern Observatory (ESO), using Target of Opportunity (ToO)
and Visitor modes, in 3 domains: optical ($400-800$\,nm) with the EMMI
instrument on the 3.5m New Technology Telescope (NTT) at La Silla, NIR
($1-2.5 \microns$) with the SOFI instrument on the NTT, and
mid-infrared (MIR, $5-20 \microns$) with the VISIR instrument on
Melipal, the 8m Unit Telescope 3 (UT3) of the Very Large Telescope
(VLT) at Paranal (Chile).  These observations include photometry and
spectroscopy on 21 {\it INTEGRAL} sources in order to identify their
counterparts and the nature of the companion star, derive the distance,
and finally characterise the presence and temperature of their
circumstellar medium.

     \subsection{IGR~J16318-4848: extreme among the obscured high-energy
sources}

IGR~J16318-4848 was the first source to be discovered by IBIS/ISGRI on
{\it INTEGRAL} on 29 January 2003 \citep{courvoisier:2003}. {\it
XMM-Newton} observations showed a strong absorption of $\nh \sim 2
\times 10^{24} \cmmoinsdeux$ \citep{matt:2003}.  The accurate
localisation by {\it XMM-Newton} allowed \cite{filliatre:2004} to
rapidly trigger ToO photometric and spectroscopic observations in
optical and NIR, leading to the discovery of the optical counterpart
and to the confirmation of the NIR one found by \cite{walter:2003}.
The extremely bright NIR source 
(Ks\,$=7.20 \mags$) exhibits an unusually strong intrinsic
absorption in the optical of $A_v = 17.4 \mags$, much stronger than
the absorption along the line of sight of $A_v = 11.4 \mags$, but
still 100 times lower than the absorption in X-rays.  This led
\cite{filliatre:2004} to suggest that the material absorbing in the
X-rays was concentrated around the compact object, while the material
absorbing in the optical/NIR was enshrouding the whole system.  The
NIR spectroscopy 
revealed an unusual
spectrum, with many strong emission lines, originating from a highly
complex and stratified circumstellar environment, of various densities
and temperatures, suggesting the presence of an envelope and strong
stellar outflow, responsible for the absorption. Only luminous
early-type stars such as supergiant sgB[e] show such extreme
environments, and \cite{filliatre:2004} concluded that IGR~J16318-4848
was an unusual HMXB.
By combining these optical and NIR data with MIR observations, and
fitting these observations with a model of a sgB[e] companion star,
\cite{rahoui:2007} showed that IGR~J16318-4848 exhibits a MIR
excess (see Figure \ref{figure:igrj16318-igrj17544}, left panel), that they
interpret as being due to the strong stellar outflow emanating from
the sgB[e] companion star.  They found that the companion star had a
temperature of T\,$=23500$\,K and radius R$_{\star} = 20.4 R_{\odot}$, and
an extra component of temperature T $=900$\,K and radius R\,$= 12
R_{\star}$, with A$_v = 17.6 \mags$. 
If we take a typical orbital period of 10 days and a
mass of the companion star of $20 \Msol$, we obtain an orbital
separation of $50 \Rsol$, smaller than the
extension of the extra
component, suggesting that this component enshrouds the whole binary system,
as would do a cocoon of gas/dust (see Figure \ref{figure:obscured-sfxt},
left panel).
In summary, IGR~J16318-4848 is an HMXB system, located at a distance
between 1 to 6 kpc, hosting a compact object (probably a neutron star)
and a sgB[e] star (it is therefore the second HMXB with a sgB[e] star,
after CI Cam; \citeauthor{clark:1999} \citeyear{clark:1999}). The most
striking facts are i. the compact object seems to be surrounded by
absorbing material and ii. the whole binary system seems to be
surrounded by a dense and absorbing circumstellar material envelope or
cocoon, made of cold gas and/or dust. This source exhibits so extreme
characteristics that it might not be fully representative of the other
obscured sources.

     \subsection{IGR~J17544-2619: archetype of the Supergiant Fast
X-ray Transients}

\begin{figure}
  \includegraphics[height=.3\textheight,angle=-90]{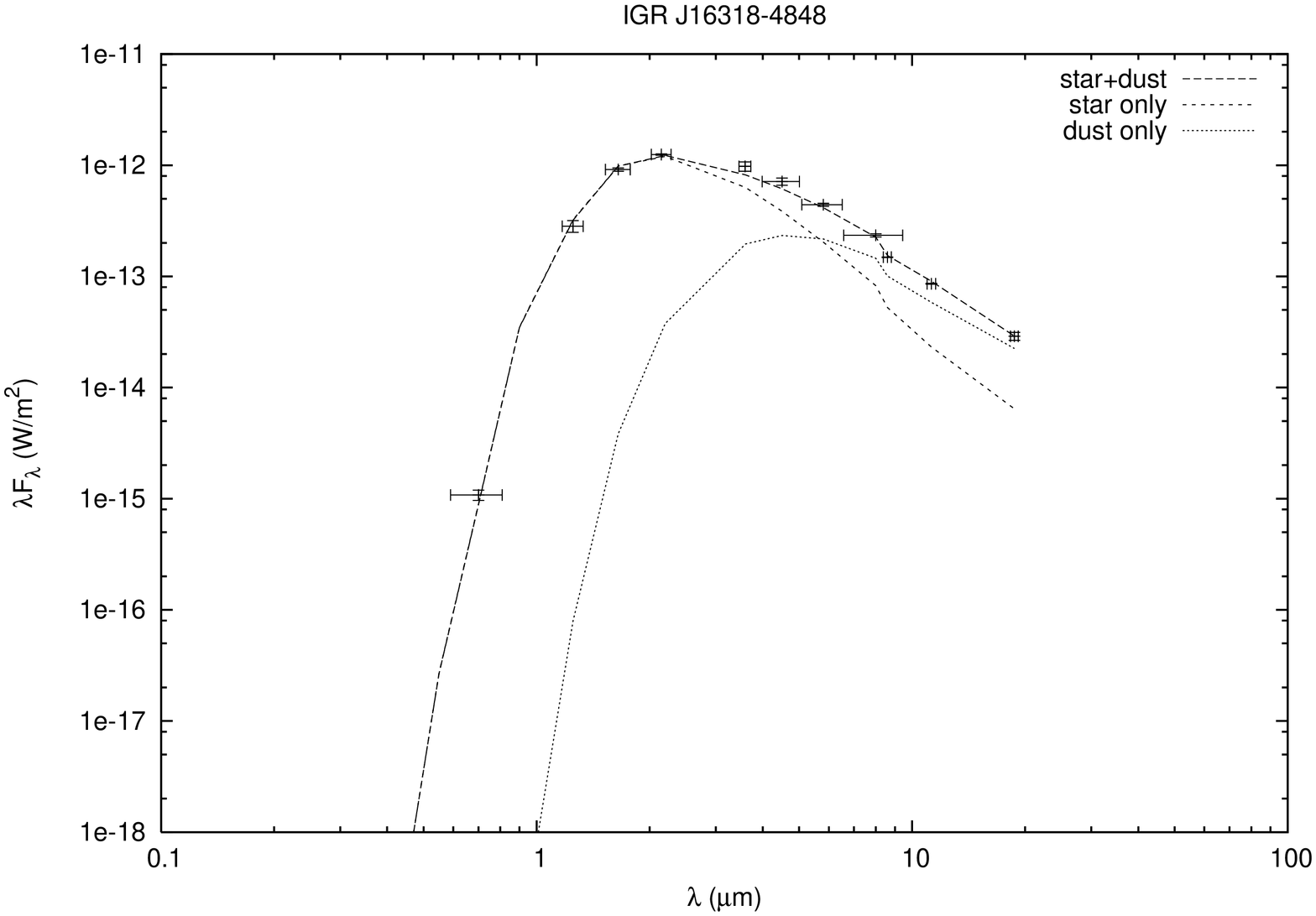}
  \includegraphics[height=.3\textheight,angle=-90]{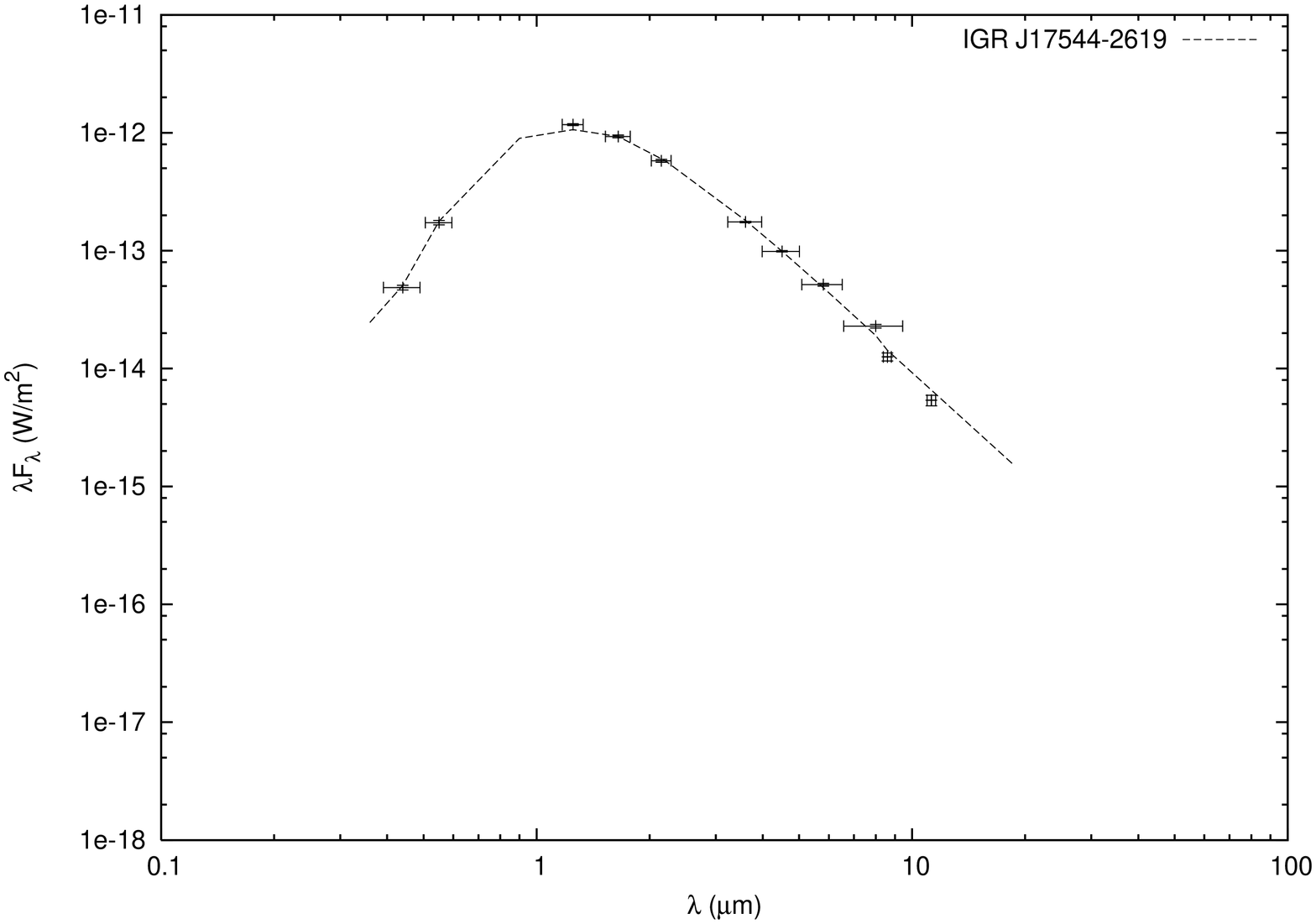}
  \caption{\label{figure:igrj16318-igrj17544} Optical to MIR SEDs of
  IGR~J16318-4848 (left) and IGR~J17544-2619 (right), including data
  from ESO/NTT, VISIR on VLT/UT3 and {\it Spitzer} \citep{rahoui:2007}.
  IGR~J16318-4848 exhibits a MIR excess, interpreted by
  \cite{rahoui:2007} as the signature of a strong stellar outflow
  coming from the sgB[e] companion star \citep{filliatre:2004}.  On the
  other hand, IGR~J17544-2619 is well fitted with only a stellar
  component corresponding to the O9Ib companion star spectral type
  \citep{pellizza:2006}.}
\end{figure}

SFXTs constitute a new class of sources identified among the recently
discovered {\it INTEGRAL} sources, whose common characteristics are:
they exhibit rapid outbursts lasting only hours, a faint quiescent
emission, their high energy spectra require a BH or NS accretor, and
they host O/B supergiant companion stars. IGR~J17544-2619, a bright
recurrent transient X-ray source discovered by {\it INTEGRAL} on 17
September 2003 \citep{sunyaev:2003b}, seems to be the archetype of
this class of sources. Observations with {\it XMM-Newton} have shown
that it exhibits a very hard X-ray spectrum, and a relatively low
intrinsic absorption ($10^{22} \cmmoinsdeux$,
\citeauthor{gonzalez-riestra:2004} \citeyear{gonzalez-riestra:2004}).
Its bursts last for hours, in-between bursts it exhibits long
quiescence periods, which can reach more than 150 days (Zurita Heras
et al. in prep.). The compact object is probably a
neutron star \citep{intzand:2005}.  \cite{pellizza:2006} managed to
get optical/NIR ToO observations only one day after the discovery of
this source. They identified a likely counterpart inside the {\it
XMM-Newton} error circle, confirmed by {\it Chandra} accurate
localization.  Spectroscopy showed that the companion star was a blue
supergiant of spectral type O9Ib, with a mass of $25-28 M_{\odot}$ and
temperature of T $\sim 31000$ K: the system is therefore an HMXB
\citep{pellizza:2006}.
\cite{rahoui:2007} combined optical, NIR and MIR observations and
showed that they could accurately fit the observations with a model of
an O9Ib star: temperature T~$=30500$~K and radius R$_{\star} = 21.9
R_{\odot}$. They derived an absorption A$_v = 5.9 \mags$ and a
distance D~$=3.9$~kpc.  The source does not exhibit any MIR excess
(see Figure \ref{figure:igrj16318-igrj17544}, right panel,
 \citeauthor{rahoui:2007} \citeyear{rahoui:2007}).
In summary, IGR~J17544-2619 is an HMXB at a distance of $\sim$4~kpc,
constituted of an O9Ib supergiant, with a mild stellar wind and a
compact object which is probably a neutron star, without any MIR excess.

     \subsection{General results on {\it INTEGRAL} sources and discussion} \label{IGRs}

\begin{figure*}
  \includegraphics[height=.34\textheight,angle=-90]{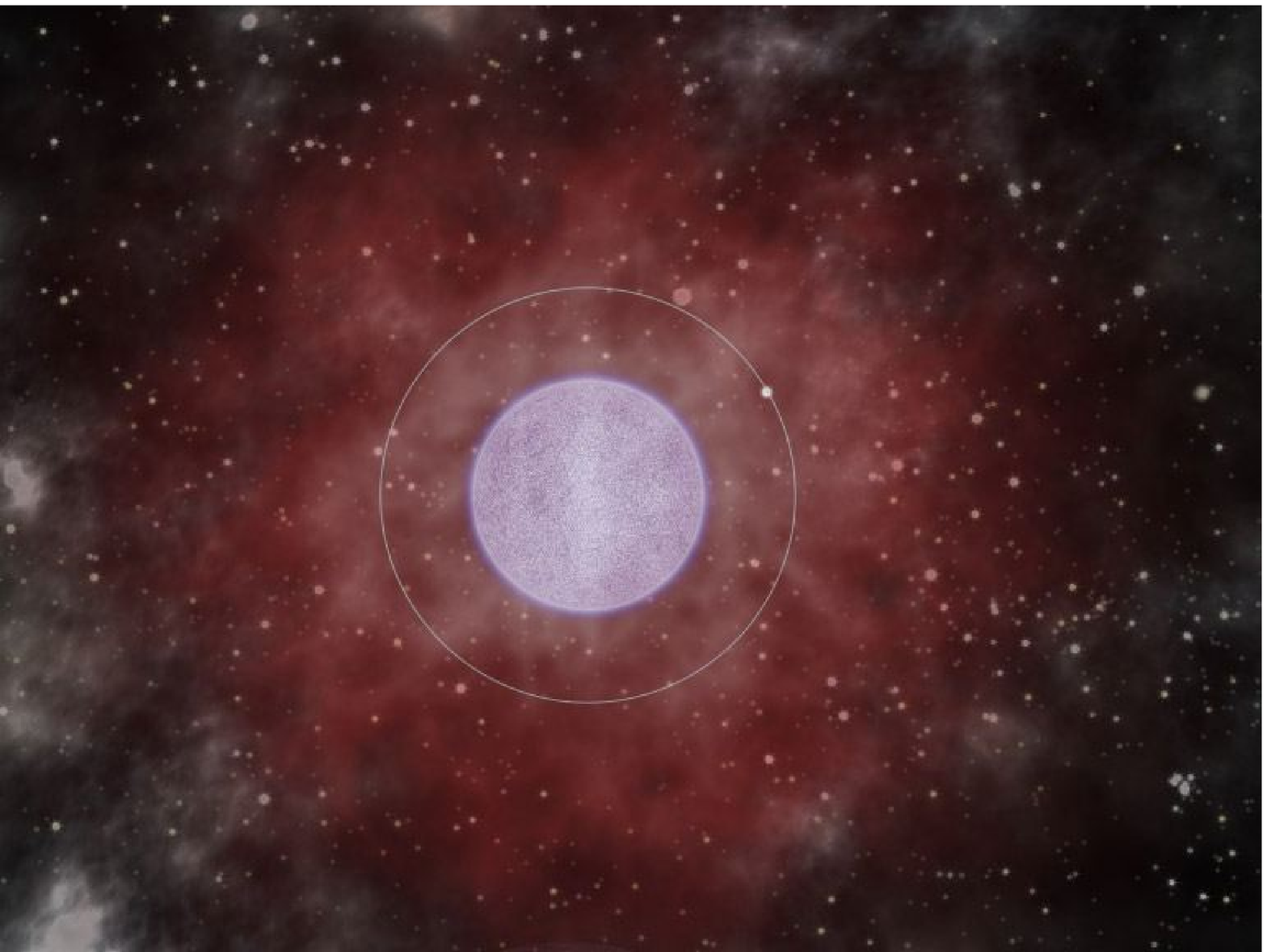}
  \includegraphics[height=.34\textheight,angle=-90]{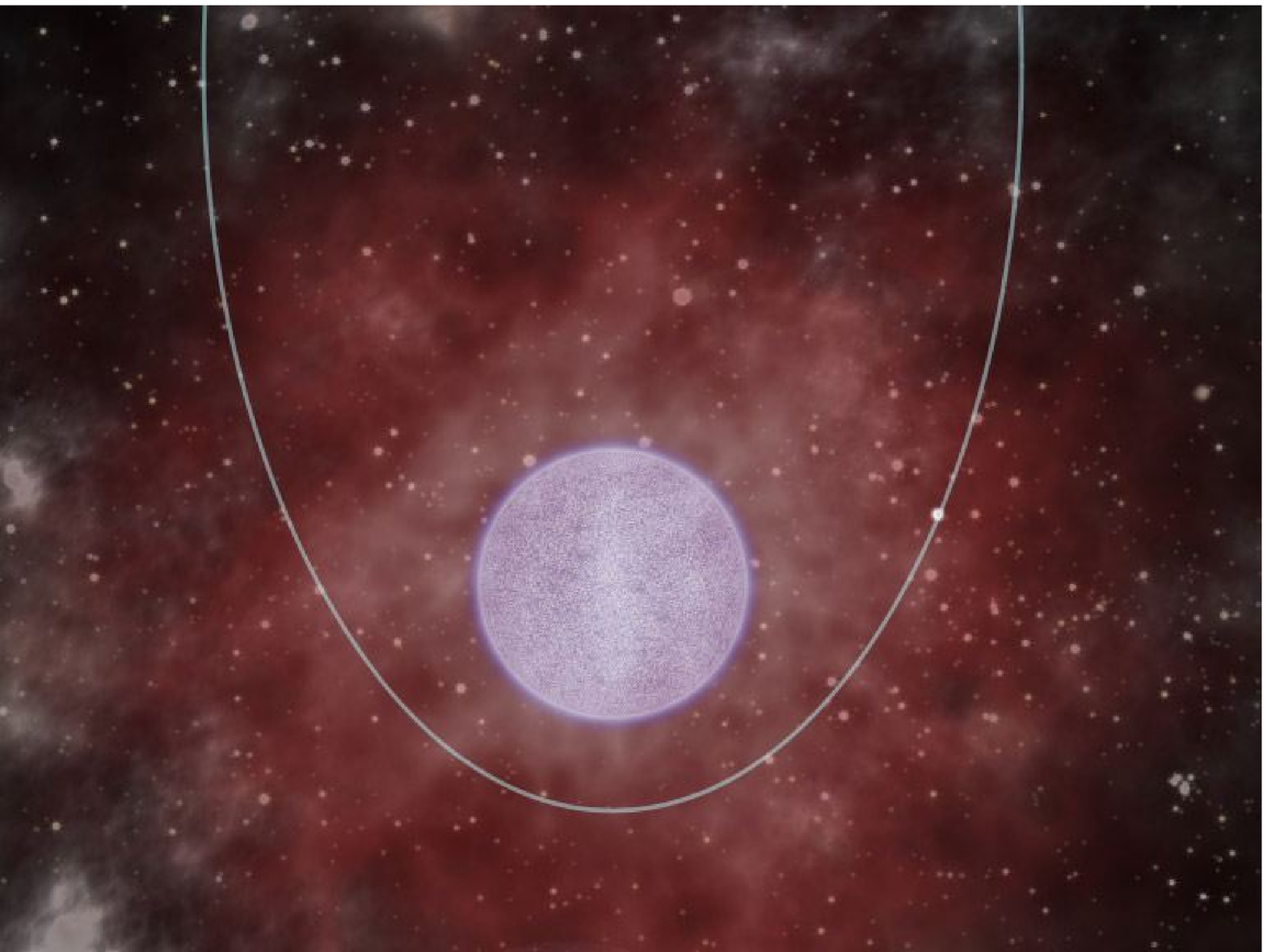}
  \caption[Scenarios illustrating both 2 types of {\it INTEGRAL}
sources]{Scenarios illustrating two possible configurations of {\it
INTEGRAL} sources, with a neutron star orbiting around a supergiant
star on a circular orbit (left image), and on an excentric orbit
(right image), accreting from the clumpy stellar wind of the
supergiant.  The accretion of matter is persistent in the case of the
obscured sources, as in the left image, where the compact object
orbits inside the cocoon of dust enshrouding the whole system. On the
other hand, the accretion is intermittent in the case of SFXTs, which
might correspond to a compact object on an excentric orbit, as in the
right image.  A 3D animation of these sources is available on the
website {\em http://www.aim.univ-paris7.fr/CHATY/Research/hidden.html.}}
  \label{figure:obscured-sfxt}
\end{figure*}

To better characterize this population, \cite{chaty:2007b} and
\cite{rahoui:2007} studied a sample of 21 {\it INTEGRAL} sources
belonging to both classes described above.
Some results are reported in Table \ref{table:results}. The optical/NIR
study, through an accurate astrometry, photometry and spectroscopy,
allowed \cite{chaty:2007b} to identify 
the counterpart, and to show that most of these systems are HMXBs,
containing massive and luminous early-type companion stars. By
combining MIR photometry, and fitting their optical--MIR spectral
energy distributions, \cite{rahoui:2007} showed that i. most of these
sources exhibit an intrinsic absorption and ii. three of them exhibit
a MIR excess, that they suggest to be due to the presence of a cocoon
of dust and/or cold gas enshrouding the whole binary system (see also
\citeauthor{chaty:2006c} \citeyear{chaty:2006c}).
Nearly all the {\it INTEGRAL} HMXBs for which both spin and orbital
periods have been measured are located in the upper part of the Corbet
diagram \citep{corbet:1986}: they are wind accretors, typical of
supergiant HMXBs, and X-ray pulsars exhibiting longer pulsation
periods and higher absorption (by a factor $\sim4$) as compared to the
average of previously known HMXBs \citep{bodaghee:2007}. This extra
absorption might be due to the presence of a cocoon of dust/cold gas 
enshrouding the whole binary system in the case of the
obscured sources: the intrinsic properties of the supergiant companion star 
could therefore explain some properties of these
sources.  However, fundamental differences exist between obscured
sources and SFXTs, which might be explained by the geometry of the
binary systems, and/or the extension of the wind/cocoon enshrouding
either the companion star or the whole system.  Indeed, obscured
sources are naturally explained by a compact object orbiting inside a
cocoon of dust and/or cold gas, while the fast X-ray behaviour of
SFXTs needs a clumpy stellar wind environment, to account for fast and
transient accretion phenomena (see Figure \ref{figure:obscured-sfxt},
left and right panels respectively, and \citeauthor{chaty:2006c}
\citeyear{chaty:2006c}).
%
These results show the existence in our Galaxy of a dominant
population of a previously rare class of high-energy binary systems:
supergiant HMXBs, some exhibiting a high intrinsic absorption
(\citeauthor{chaty:2007b} \citeyear{chaty:2007b};
\citeauthor{rahoui:2007} \citeyear{rahoui:2007}).  A careful study of
this population, recently revealed by {\it INTEGRAL}, will provide a
better understanding of the formation and evolution of short-living
HMXBs.  Furthermore, stellar population models will henceforth have to
take these objects into account, to assess a realistic number of
high-energy binary systems in our Galaxy. Our final word is that only
a multiwavelength study can allow to reveal the nature of the
obscured high-energy sources.

\begin{table*}
  \begin{center}
    \caption{Results on the sample of {\it INTEGRAL} sources, more
details are given in \cite{chaty:2007b}. We indicate respectively the
name of the sources, the region of the Galaxy in the direction of
which they are located, their spin and orbital period,
the interstellar, optical-IR and X-ray derived column density
respectively (in units of $10^{22}\cmmoinsdeux$), their spectral type,
nature and reference.
Type abbreviations: 
AGN = Active Galactic Nucleus,
B = Burster,
BHC = Black Hole Candidate,
CV = Cataclysmic Variable,
D = Dipping source,
H = High Mass X-ray Binary system, 
IP = Intermediate polar, 
L = Low Mass X-ray Binary,
O = Obscured source,
P = Persistent source,
S = Supergiant Fast X-ray Transient,
T: Transient source,
XP: X-ray Pulsar.
Reference are:
c: \cite{chaty:2007b},
co: \cite{combi:2006},
f: \cite{filliatre:2004},
h: \cite{hannikainen:2007},
m1: \cite{masetti:2004}
m2: \cite{masetti:2006}
n1: \cite{negueruela:2005},
n2: \cite{negueruela:2006a},
n3: \cite{nespoli:2007},
p: \cite{pellizza:2006},
t: \cite{tomsick:2006a}.
}
\label{table:results}
\vspace{1em}
    \renewcommand{\arraystretch}{1.2}
    \begin{tabular}{cccccccccc} 
\hline
Source & Reg & P$_s$(s) & P$_o$(d) & $Nh_{is}$ & $Nh_{IR}$ & $Nh_X$ & SpT & Type & Ref \\
\hline
IGR J16167-4957 & No & & & 2.2 & 0.23 & 0.5 & A0 & CV/IP & t,m2 \\
\hline 
IGR J16195-4945 & No & & & 2.18 & 2.9 & 7 & OB & H?/S?/O & t \\
\hline
IGR J16207-5129 & No & & & 1.73 & 2.0 & 3.7 & BOI & H/O & t,m2 \\
\hline
IGR J16318-4848 & No & & & 2.06 & 3.3 & 200 & sgB[e] & H/O/P & f \\
\hline
IGR J16320-4751 & No & 1250 & 8.96(1) & 2.14 & 6.6 & 21 & sgOB & H/XP/T/O & c \\
\hline
IGR J16358-4726 & No & 5880 & & 2.20 & 3.3 & 33 & sgB[e]? & H/XP/T/O & c \\
\hline
IGR J16393-4643 & No & 912 & 3.6875(6) & 2.19 & 2.19 & 24.98 & BIV-V? & H/XP/T & c \\
\hline 
IGR J16418-4532 & No & 1246 & 3.753(4) & 1.88 & 2.7 & 10 & sgOB? & H/XP/S & c \\
\hline
IGR J16465-4507& No & 228 & & 2.12 & 1.1 & 60 & B0.5I & H/S & n1 \\
\hline
IGR J16479-4514 & No & & & 2.14 & 3.4 & 7.7 & sgOB & H/S? & c \\
\hline
IGR J16558-5203 & - & - & - & - & - & - & Sey1.2 & AGN & m2 \\
\hline 
IGR J17091-3624 & GC & & & 0.77 & 1.03 & 1.0 & L & L/BHC & c \\
\hline 
IGR J17195-4100 & GC & & & 0.77 & & 0.08 & & CV/IP & t,m2 \\
\hline 
IGR J17252-3616 & GC & 413 & 9.74(4) & 1.56 & 3.8 & 15 & sgOB & H/XP/O & c \\
\hline
IGR J17391-3021 & GC & & & 1.37 & 1.7 & 29.98 & O8Iab(f) & H/S/O & n2 \\
\hline
IGR J17544-2619 & GC & & 165? & 1.44 & 1.1 & 1.4 & O9Ib & S & p \\ 
\hline
IGR J17597-2201 & GC & & & 1.17 & 2.84 & 4.50 & L & L/B/D/P & c \\
\hline
IGR J18027-1455 & - & - & - & - & - & - & Sey1 & AGN & m1,co \\
\hline 
IGR J18027-2016 & GC & 139 & 4.5696(9) & 1.04 & 1.53 & 9.05 & sgOB & H/XP/T & c \\
\hline
IGR J18483-0311 & GC & 21.05 & 18.55 & 1.62 & 2.45 & 27.69 & BeV? & H/XP & c \\
\hline 
IGR J19140+0951 &    & & 13.558(4) & 1.68 & 2.9 & 6 & sgB0.5I & H/O & n3,h \\
\hline
\end{tabular}
  \end{center}
\end{table*}

\begin{acknowledgements}
SC would like to thank the organisers for their invitation to report on
these exciting results on newly discovered {\it INTEGRAL} sources, 
and for organising an interesting workshop, in a nice place,
fruitful to arise scientific discussions and new ideas. 
SC is grateful to Juan Antonio Zurita Heras for a careful rereading
of the manuscript, and to an anonymous referee for useful comments.
\end{acknowledgements}




\bigskip
\bigskip
\noindent {\bf DISCUSSION}

\bigskip
\noindent {\bf NIELS LUND:} Are all the SFXT sources heavily absorbed?
The animation seems to indicate a connection between the heavily
absorbed sources and the SFXT sources.

\bigskip
\noindent {\bf SYLVAIN CHATY:} No, the SFXTs are not all heavily
absorbed.  The animation shows two possible configurations for
INTEGRAL sources.  The first one, where the neutron star is on a
circular orbit inside the cocoon of dust, well corresponds to the
obscured sources. The second one, where the neutron star on an
excentric orbit only periodically crosses the clumpy environment of
its companion supergiant star, could explain some properties of the
SFXTs.

\label{lastpage}

\end{document}